\documentclass[aps,12pt,preprint,nofootinbib,onecolumn,preprintnumbers,superscriptaddress,noshowpacs]{revtex4}

\usepackage{amssymb, amsfonts, amsmath, bm}
\usepackage%[dvipdfmx]
{graphicx}
\usepackage{color}
\usepackage{float}
\usepackage{ascmac}

\usepackage{ulem}
\usepackage%[dvipdfmx]
{hyperref}
\hypersetup{% 
 setpagesize=false,
 bookmarksnumbered=true,%
 bookmarksopen=true,%
 colorlinks=true,%
 linkcolor=blue,%
 citecolor=red}
\begin{document}
\small
\rightline{
\baselineskip16pt\rm\vbox to 50pt{
     \hbox{RUP-19-11}
     \hbox{OCU-PHYS-500}
     \hbox{AP-GR-154}
     \hbox{NITEP-16}
}
}

\title{{\Large
Observability of spherical photon orbits\\
in near-extremal Kerr black holes
}
}
\author{{\large Takahisa Igata}}
\email{igata@rikkyo.ac.jp}
\affiliation{Department of Physics, Rikkyo University, Toshima, Tokyo 171-8501, Japan}
\author{{\large Hideki Ishihara}}
\email{ishihara@sci.osaka-cu.ac.jp}
\affiliation{Department of Mathematics and Physics,
Graduate School of Science, Osaka City University,
Osaka 558-8585, Japan}
\author{{\large Yu Yasunishi}}
%\email{yasunishi@sci.osaka-cu.ac.jp}
\affiliation{Department of Mathematics and Physics,
Graduate School of Science, Osaka City University,
Osaka 558-8585, Japan}

\date{\today}

\begin{abstract}
We investigate the spherical photon orbits 
in near-extremal Kerr spacetimes. 
We show that the spherical photon orbits 
with impact parameters
in a finite range 
converge on the event horizon. 
Furthermore, we 
demonstrate that the Weyl curvature 
near the horizon does not generate 
the shear of a congruence of such light rays. 
Because of this property, 
a series of images produced by the light
orbiting around 
a near-extremal Kerr black hole several times can be observable. 
\end{abstract}

\maketitle

%%%%%
\section{Introduction}
%%%%%

A large number of gravitational lens systems have been observed 
in our Universe~\cite{Wambsganss:1998gg}. 
In the case of the gravitational lensing by a black hole, 
it would be possible that 
we observe a series of images produced by a direct light, 
light orbiting around the black hole once, twice, and so on. 
The series of images are related to the unstable photon orbits 
with constant radii 
around the black hole, the so-called spherical photon orbits~\cite{Teo}. 
The photons escaping from the spherical photon orbits toward an 
observer make the series of images.

Very recently, the Event Horizon Telescope Collaboration reported 
observations of a bright emission ring in the central 
region of M87~\cite{Akiyama:2019cqa, Akiyama:2019brx, 
Akiyama:2019sww,Akiyama:2019bqs, Akiyama:2019fyp, Akiyama:2019eap} 
by using a global very long baseline 
interferometry array. 
It is interpreted that the observed ring is produced by photons orbiting 
the supermassive black hole sitting in the center of M87.

There are a lot of black hole candidates, 
and many of them are thought to be 
rapidly rotating black holes~\cite{Reynolds:2013qqa}. 
Therefore, it is important to investigate 
the gravitational lensing by rapidly rotating black holes. 
It is well known that 
the radius of the prograde circular photon orbit 
on the equatorial plane of a Kerr black hole 
approaches the horizon radius in the extremely rotating limit. 
If we can observe the images related 
to such photon orbits, we obtain information on 
the near-horizon
geometry of the extremal Kerr black hole.

In a slowly rotating black hole case, 
it would be hard to observe the images produced by light orbiting around the black hole 
because the brightness of the images 
decreases as the light ray winds. 
The Weyl curvature around the black hole 
generates the shear of a congruence of the light rays, 
and the shear induces the expansion of the congruence. 
Hence the brightness of the images decreases exponentially as 
the number of windings increases.

In an extremely rotating black hole case, 
it is known that the near-horizon geometry 
with long throat structure admits 
enhanced symmetry~\cite{Bardeen:1972fi, Bardeen:1999px}.
Recently, in the context of the Kerr/CFT correspondence~\cite{Guica:2008mu, Compere:2012jk}, 
the near-horizon region of the near-extremal Kerr black hole 
has been an interesting area that provides new phenomena. 
Then, we suppose that 
spherical photon orbits that exist in the near-horizon region 
in the near-extremal Kerr black hole have different properties from the ones 
in a slowly rotating 
black hole~\cite{Gralla:2017ufe, Lupsasca:2017exc}. 
We show, in this article, that the Weyl curvature 
does not generate the shear of a congruence 
of the spherical photon orbits 
near the horizon of near-extremal Kerr black holes.

%%%%%
\section{Spherical photon orbits in Kerr spacetimes} 
%%%%%

The Kerr metric in the Boyer--Lindquist 
coordinates $(t,r,\theta,\varphi)$ is given by
\begin{align}
	&g_{\mu\nu} \mathrm{d}x^\mu \mathrm{d}x^\nu
	=-\frac{\Delta \Sigma}{A} \mathrm{d}t^2
		+\frac{\Sigma}{\Delta} \mathrm{d}r^2
		+\Sigma \:\!\mathrm{d}\theta^2
	+\frac{A}{\Sigma}\sin^2\theta \!
	\left[\mathrm{d}\varphi-\frac{2Mar}{A} \mathrm{d}t\right]^2,
\\
	&\Sigma=r^2+a^2\cos^2\theta,
\ \ 
	\Delta=r^2+a^2-2Mr,
\ \ 
	A=\left(r^2+a^2\right)^2-a^2\Delta \sin^2\theta.
\end{align}
When $| a |\leq M$, the metric describes 
a rotating black hole with mass $M$ and 
specific angular momentum $a$. 
The black hole spacetime has 
the event horizon at $r=r_+:=M+\sqrt{M^2-a^2}$ and 
the inner horizon at $r=r_-:=M-\sqrt{M^2-a^2}$. 
The event horizon is generated by the null tangent vector field
\begin{align}
\label{eq:chi}
	\chi^a=(\partial/\partial t)^a
	+\Omega_{\textrm{h}} (\partial/\partial \varphi)^a,
\end{align}
where $\Omega_{\textrm{h}}=a/(r_+^2+a^2)$ is 
the angular velocity of the event horizon. 
We use units in which $M=1$ in what follows.

Let $k^a$ be a tangent vector to the null geodesics 
parametrized by an affine parameter~$\lambda$ 
in the Kerr spacetime. 
According to the time translation symmetry and axisymmetry, 
$E=-k_t$ and $L=k_\varphi$ are constants of motion.
In addition, we have a constant of motion~\cite{Carter:1968rr},
\begin{align}
	Q=K_{ab} k^a k^b-(L-aE)^2,
\end{align}
where $K_{ab}$ is the Killing tensor defined by
\begin{align}
\label{eq:KT}
	K_{ab}=\Sigma^2 (\mathrm{d}\theta)_a (\mathrm{d}\theta)_b
		+\sin^2\theta \left[ 
	\left(r^2+a^2\right)(\mathrm{d}\varphi)_a-a (\mathrm{d}t)_a
	 \right]\left[ 
	\left(r^2+a^2\right)(\mathrm{d}\varphi)_b-a (\mathrm{d}t)_b
	 \right]-a^2\cos^2\theta g_{ab}.
\end{align}
Introducing the dimensionless impact parameters 
\begin{align}
	b=\frac{L}{E}, \ \ q=\frac{Q}{E^2},
\end{align}
for nonvanishing $E$, 
since $K_{ab}k^ak^b \geq 0$, we have the inequality 
\begin{align}
\label{eq:lowerq}
	q+(b-a)^2\geq 0.
\end{align}

In terms of the parameters $b$ and $q$, the null geodesic equations are 
\begin{align}
	&k^t=\dot{t}=\frac{1}{\Sigma}\left[
		a \left(b-a\sin ^2\theta \right)
		+\frac{r^2+a^2}{\Delta} \left[r^2+a(a-b)\right]
		\right],
		\ \ 
	k^r=\dot{r}=\frac{\sigma_r}{\Sigma}\sqrt{-V}, 
\cr
	&k^\theta =\dot{\theta}=\frac{\sigma_\theta}{\Sigma}\sqrt{-U},
\ \ 
	k^\phi =\dot{\varphi}
	=\frac{1}{\Sigma}\left[ \frac{b}{\sin^2\theta}-a
	+\frac{a}{\Delta} \left[r^2+a(a-b)\right]
		 \right],
\label{eq:eom}
\end{align}
where $\sigma_r$, $\sigma_\theta=\pm1$, the dots 
denote derivatives with respect to $\lambda$, and functions $V, U$ are given by 
\begin{align}
	&V=\Delta \left[ q+(b-a)^2 \right]-\left[r^2+a(a-b)\right]^2, 
\\
&
\label{eq:Theta}
	U=\cos^2\theta\left(\frac{b^2}{\sin^2\theta}-a^2 \right)-q. 
\end{align}
From the equation of motion for $\theta$ in Eq.~\eqref{eq:eom}, 
$U\leq 0$ should hold, 
and then, the allowed region of $\theta$ is 
classified into the following three cases: 
\begin{align}
\label{case1}
(\textrm{i}) 
	&\quad \text{If} \quad q>0,\quad 
	\text{then} ~~ |\cos \theta | \leq u_+, 
\\[1mm]
\label{case2}
(\textrm{ii}) 
	&\quad \text{If} \quad q=0 ~~ \text{and}~~ b^2>a^2, \quad
	\text{then}~~ \theta=\pi/2,
\\[1mm]
\label{case3}
(\textrm{iii}) 
	& \quad \text{If}\quad q\leq 0 ~~\text{and}~~ 
	b^2\leq \left(a-\sqrt{-q} \right)^2, \quad
	\text{then}~~ u_- \!\leq |\cos \theta | \leq u_+,
\end{align}
where 
\begin{align}
	u_\pm=\frac{1}{\sqrt{2} a}\!\left[ a^2-b^2-q\pm\sqrt{(a^2-b^2-q)^2+4 a^2 q} \right]^{1/2}. 
\end{align}
In a special case $(b, q)=(a \sin^2\theta_0, -a^2\cos^4\theta_0)$ in (iii), 
where $\theta_0$ is a constant,
$k^a$ is identified with the principal null vectors in the Kerr spacetime: 
\begin{align}
N^a_{\pm}=\frac{r^2+a^2}{\Delta}
	\left(\partial/\partial t\right)^a
		\pm \left(\partial/\partial r\right)^a
		+\frac{a}{\Delta}\left(\partial/\partial \varphi\right)^a.
\end{align}

Hereafter, we focus on the spherical photon orbits~\cite{Teo},
$\dot{r}=0$ and $\ddot{r}=0$, 
in the range $r_+<r$ or 
$0< r<r_-$.
Then, the radial equation in Eq.~\eqref{eq:eom} leads to the equations
\begin{align}
\label{eq:V0dV0}
	V=0, \quad 
	\frac{\mathrm{d}V}{\mathrm{d}r}=0.
\end{align}
Solving coupled algebraic equations~\eqref{eq:V0dV0} for $b$ and $q$, 
we obtain 
\begin{align}
\label{eq:bqsol1}
	b&=\frac{2(1-a^2)}{a(r-1)}-\frac{\left(r-1\right)^2}{a}
	+\frac{3-a^2}{a},
\\
\label{eq:bqsol2}
	q
	&=-\frac{4(1-a^2)}{a^2(r-1)^2}-\frac{12(1-a^2)}{a^2(r-1)}
\cr
	&
	\quad	
	+\frac{3}{a^2}(4a^2-3) 
	 +\frac{4}{a^2}(1+a^2)(r-1)+\frac{6}{a^2}(r-1)^2
		-\frac{1}{a^2}(r-1)^4. 
\end{align}

%%%%%
\begin{figure}[!h]
\centering
\includegraphics[width=15cm,clip]{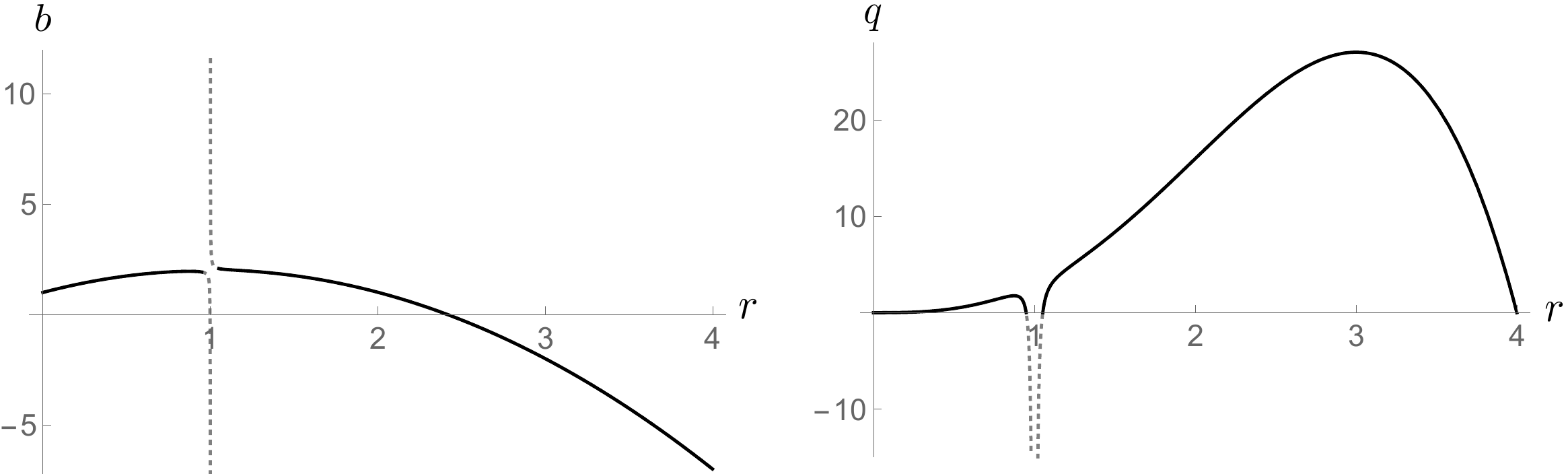}
 \caption{
Plots of $b$ (left) and $q$ (right) 
as functions of $r$ for $a=0.999$. 
 The solid lines show the values of $b$ and $q$ 
 for each radius of spherical photon orbits 
in the range $0<r\leq r_3$ and $r_1\leq r \leq r_2$. 
}
\label{fig:1}
\bigskip
\end{figure}
%%%%%

Figure~\ref{fig:1} shows $b$ and $q$ 
as functions of the radius of spherical photon orbits, $r$, 
in a near-extremal Kerr black hole. 
Since Eqs.~\eqref{eq:bqsol1} and \eqref{eq:bqsol2} 
lead to the inequality $b^2>a^2-q$, 
then any spherical photon orbit does not fall in the case of (iii), 
and therefore $q \geq 0$. 
It implies that the spherical photon orbits appear in the range
\begin{align}
\label{eq:ranger0}
	r_+<r_1\leq r\leq r_2,
\quad
	0<r\leq r_3<r_-, 
\end{align}
where
\begin{align}
\label{eq:r1}
	&r_1=2+2\cos \left[ \frac{2}{3} \mathrm{Arccos}(a)
	-\frac{2\pi}{3} \right],
\\
\label{eq:r2}
	&r_2=2+2\cos\left[ \frac{2}{3}\mathrm{Arccos}(a) \right],
\\
\label{eq:r3}
	&r_3=2+2 \cos\left[ \frac{2}{3} \mathrm{Arccos}(a)
	+\frac{2\pi}{3} \right]
\end{align}
are solutions to the equation $q=0$. 
In the extremal limit, $a\to 1$, we see that $r_1 \to r_+$ and $r_3 \to r_-$, 
and then, on the equatorial plane, there exists a circular photon 
orbit that approaches the event horizon.

%%%%%
\section{Spherical photon orbits that approach the horizon in the extremal limit}
%%%%%

\begin{figure}[!h]%[htbp]
\centering
 \includegraphics[width=6.5cm,clip]{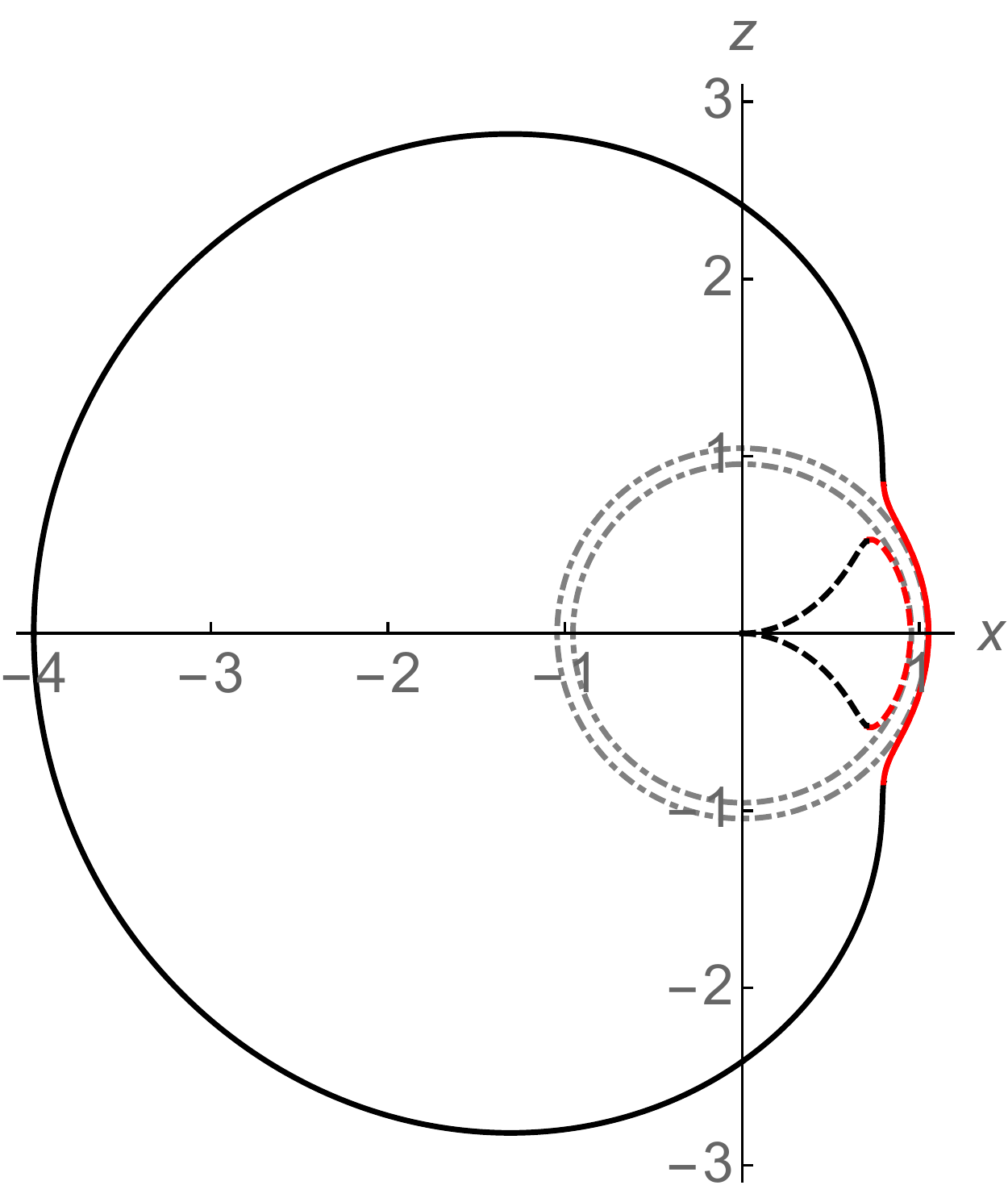}
 \caption{
For the spherical photon orbits in a 
near-extremal Kerr spacetime with $a=0.999$, 
the relation of $r$ and $\theta_\text{min/max}$ 
 in the $x$-$z$ plane is shown, 
where 
$(x, z)
=(\mathrm{sgn}(b)~ r\sin \theta_\text{min/max}, 
r \cos \theta_\text{min/max})$. 
 The black solid curve corresponds to the cardioid class 
 outside the outer horizon, 
 and the black dashed curve does that inside the inner horizon. 
 The red solid curve and the red dashed curve 
 denote the horizon class outside the outer horizon 
and inside the inner horizon, respectively. 
 The gray dashed-dotted curves show the outer horizon $r=r_+$, 
 and the inner one $r=r_-$. 
}
 \label{fig:2}
\bigskip
\end{figure}

For a spherical photon orbit with a radius $r$, 
a set of impact parameters 
$b$ and $q$ are given by 
Eqs.~\eqref{eq:bqsol1} and \eqref{eq:bqsol2}. 
Then, the polar angle $\theta$ of the orbit varies in the range 
$\theta_\text{min}\leq \theta \leq \theta_\text{max}$, 
where $\theta_\text{min/max}$, are given by 
\begin{align}
	\cos{\theta_\text{min/max}}=u_+.
\end{align}
With respect to $r$ and $\theta_\text{min/max}$, we 
plot $(x, z)$ defined by
\begin{align}
	x:= \mathrm{sgn}(b)~ r \sin{\theta_\text{min/max}}, 
	\quad 
	z:= r \cos{\theta_\text{min/max}} 
\end{align}
for the spherical photon orbits in a near-extremal black hole 
in Fig.~\ref{fig:2}. 
There are two closed curves: one is outside the outer horizon, 
and the other is inside the inner horizon. 
In the extremal limit, $a\to 1$, the curves 
in Fig.~\ref{fig:2} converge 
to limit curves consisting of a piece of the circle, $r=1$, 
which denotes the horizon and the modified cardioid defined by%
\footnote{
The last term in the right-hand side of Eq.~\eqref{mod_cardioid}
modifies a standard cardioid (see also Ref.~\cite{Tsupko:2017rdo}). 
}
\begin{align}
	r=1 \pm \sin \theta+\sqrt{2 (1\pm \sin \theta)}.
\label{mod_cardioid}
\end{align}
We can classify the spherical photon orbits into two types: 
the orbits that approach the unit circle, and the orbits that 
approach the modified cardioid 
in the 
$r$-$\theta_\text{min/max}$ plot as $a\to 1$. 
We call the former the ``horizon class"
and the latter the ``cardioid class". 
A part of the curve outside the horizon makes the cardioid 
together with a part of the curve inside the horizon. 
There are the spherical photon orbits 
of the horizon class both outside the outer horizon 
and inside the inner horizon. 
The horizon class and the cardioid class outside the event horizon 
are joined at the radius $r_{\mathrm{cr}}^+:=1+\sqrt[3]{2(1-a^2)}$, while 
those inside the inner horizon are joined at 
the radius $r_{\mathrm{cr}}^-:=1-\sqrt[3]{1-a^2}$.
We should note that the horizon class only 
appears for $a\geq 1/\sqrt{2}$.

%%%%%

Here, we concentrate on the spherical photon orbits 
of the horizon class 
in the extremal limit, $a\to 1$. 
We introduce two small parameters $\epsilon$ and $\delta$ defined by
\begin{align}
	\epsilon:=1-a, \quad 
	\delta:=r-1,
\end{align}
where $r$ denotes the radius of spherical photon orbits. 
The radii $r_1$ and $r_3$ in Eqs.~\eqref{eq:r1} 
and \eqref{eq:r3} are expanded 
by $\epsilon$ as 
\begin{align}
	r_1=1+\sqrt{\frac{8}{3}} \epsilon^{1/2}+O(\epsilon) 
	\quad \text{and}\quad 
	r_3=1-\sqrt{\frac{8}{3}} \epsilon^{1/2}+O(\epsilon),
\end{align}
respectively. These, together with Eq.~\eqref{eq:ranger0}, 
imply that 
\begin{align}
\label{eq:order_epsilon_delta}
	\epsilon \ll 
	\sqrt{\frac{8}{3}} \epsilon^{1/2}\leq 
	| \delta | \ll 1.
\end{align}
Then, we can expand $b$ and $q$ 
in terms of $\epsilon$ and $\delta$ as 
\begin{align}
	b\simeq 2+\frac{4 \epsilon}{\delta} +4 \epsilon-\delta^2,
\quad
\label{eq:q}
	q\simeq 3 -\frac{8 \epsilon}{\delta^2}+8 \delta-\frac{24 \epsilon}{\delta}.
\end{align}

If we take the limits $\epsilon\to 0$ and $| \delta | \to 0$ 
under the condition
\begin{align}
	\epsilon \ll \sqrt{\frac{8}{3}} \epsilon^{1/2}=| \delta | \ll 1,
\end{align}
then we have 
\begin{align}
	b\to 2, \quad q\to 0.
\end{align}
On the other hand, if we take 
the limits $\epsilon \to 0$ and $|\delta| \to 0$ 
under the condition
\begin{align}
	\epsilon \ll \epsilon^{1/2} \ll | \delta | \ll 1, 
\end{align}
then we have 
\begin{align}
	b\to 2, \ q\to 3. 
\end{align}
Hence, for the spherical photon orbits 
that approach the horizon, $\delta \to 0$, in the 
extremal limit, $\epsilon \to 0$, 
the parameter $b$ approaches 2, and the parameter $q$ 
takes a value in the range $ 0 \leq q \leq 3 $. 
 
In the extremal limit, $a\to 1$, $\theta_\text{min/max}$ 
of the spherical photon orbits of the horizon class 
is given by 
\begin{align}
	|\cos \theta_\text{min/max} | 
	= \frac{1}{\sqrt{2}}\left[ \sqrt{(q+1)(q+9)}
	-q-3 \right]^{1/2}
	\leq \sqrt{2\sqrt{3}-3},
\end{align}
where the inequality is evaluated by $q=3$. 
This value was also found by 
the analysis of the near horizon 
of the extremal Kerr geometry in Refs.~\cite{
Bardeen:1973tla, 
AlZahrani:2010qb, Hod:2012ax, Porfyriadis:2016gwb}
and Ref.~\cite{Harada:2011xz} in a different context. 

\begin{figure}[!h]%[htbp]
\centering
\includegraphics[width=8.5cm,clip]{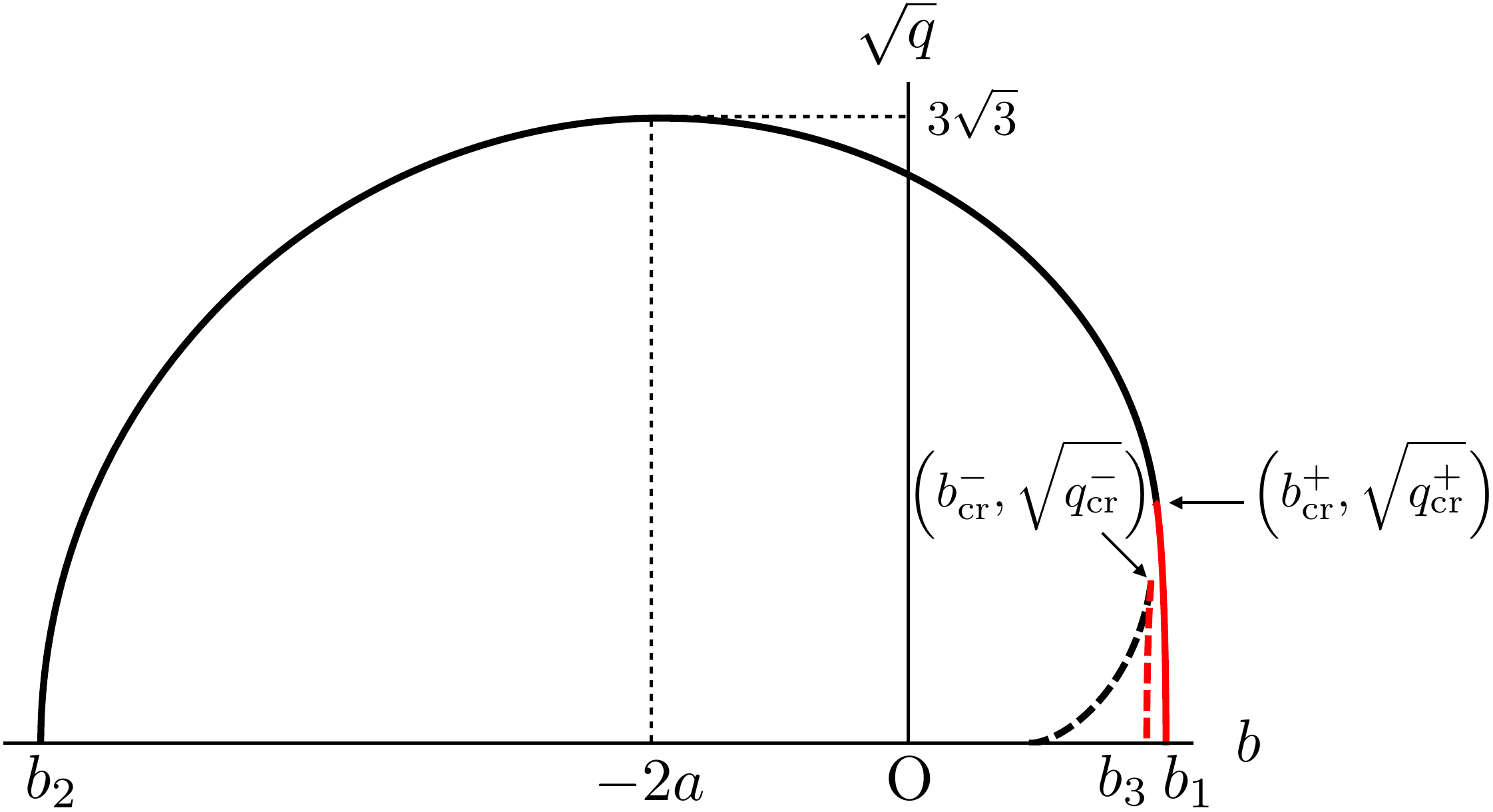}
 \caption{
The relation of $b$ and $\sqrt{q}$ is shown 
for spherical photon orbits in the case of $a=0.999$. 
The solid curve corresponds to the spherical photon orbits 
in the range $r_+ < r_1\leq r\leq r_2$, 
and the dashed curve does to the ones in $0<r\leq r_3 < r_-$. 
The photon orbits with $(b, \sqrt{q})=(b_1,0),~(b_2,0)$ are 
the prograde circular orbit at $r=r_1$ and 
the retrograde circular orbit at $r=r_2$ on the equatorial plane, respectively. 
The photon orbit with $(b_3,0)$ is 
the prograde circular orbit inside the inner horizon at $r=r_3$. 
In the limit $a\to 1$, the black solid curve and 
the black broken curve converge to a cardioid,
and the red solid curve and the red broken curve do to a straight segment. 
}
\bigskip
 \label{fig:3}
\end{figure}

We can find another cardioid 
for the spherical photon orbits 
in the relation between the impact parameters given by
Eqs.~\eqref{eq:bqsol1} and \eqref{eq:bqsol2} 
as curves in the 
$b$-$\sqrt{q}$~plane
as shown in Fig.~\ref{fig:3}. 
In the extremal limit, $a\to 1$, the curves converge to the cardioid 
that is expressed by using a parameter $\psi$ as
\begin{align}
	b-1=4\left(1-\cos \psi\right)\cos\psi. 
\cr
	\sqrt{q}=4\left(1-\cos \psi\right)\sin\psi,
\end{align}
and the straight segment that connects 
$(b, \sqrt{q})=(2, \sqrt{3})$ and $(2, 0)$.  
The horizon class corresponds to the straight segment, while 
the cardioid class does to the cardioid in the 
$b$-$\sqrt{q}$~plane.
On the curves in Fig.~\ref{fig:3}, 
the horizon class and the cardioid class are 
joined at critical points $b_\text{cr}^\pm, q_\text{cr}^\pm$ 
defined by 
$b_\text{cr}^+:=3/a-a$, 
$q_\text{cr}^+:=6 \left[ 2+\sqrt[3]{2(1-a^2)} \right]- 9/a^2$, 
and 
$b_\text{cr}^-:=(3/a)\left[ 1-\sqrt[3]{(1-a^2)^2} \right]-a,$ 
$q_\text{cr}^+:=(3/a^2)\left[ 1-\sqrt[3]{1-a^2} \right]^4$, respectively.

%%%%%%
%%%%%

%%%%%
\section{Weyl curvature near the horizon of a near-extremal Kerr spacetime}
%%%%%

We consider a property of the Weyl curvature 
near the horizon 
of the near-extremal Kerr geometry. 
To introduce a parallelly propagated tetrad along 
a null geodesic~\cite{Kubiznak:2008zs}, 
we use the conformal Killing--Yano 2-form
\begin{align}
	h=r\left( \mathrm{d}t-a \sin^2\theta \mathrm{d}\varphi \right)
	\wedge \mathrm{d}r 
		+a\cos \theta \sin\theta \left[ a \mathrm{d}t-\left(r^2+a^2\right) \mathrm{d}\varphi
		 \right]
		 \wedge \mathrm{d}\theta
\end{align}
and the Killing--Yano 2-form 
\begin{align}
	f=a \cos\theta \left( \mathrm{d}t-a \sin^2\theta \mathrm{d}\varphi \right)
	\wedge \mathrm{d}r
		- r\sin \theta \left[ a \mathrm{d}t-\left(r^2+a^2\right) \mathrm{d}\varphi \right]
		\wedge \mathrm{d}\theta.
\end{align}
These forms yield parallelly propagated unit spacelike vectors orthonormal to $k^a$, 
\begin{align}
\label{eq:m}
	m^a=\frac{k^b h_{b}{}^{a}-\lambda\left(\xi_c k^c\right)k^a}{\sqrt{C_{cd} k^c k^d}},
\quad
	n^a= \frac{k^b f_{b}{}^{a}}{\sqrt{K_{cd} k^ck^d}},
\end{align}
and a null vector external to $k^a$,
\begin{align}
	l^a=\frac{m^b h_{b}{}^{a}}{\sqrt{C_{cd} k^c k^d}}
		+\frac{C_b{}^dC_{dc} k^b k^c
		+\lambda^2\left(\xi_e k^e\right)^2C_{cd} k^c k^d }{2 \left(C_{cd} k^c k^d\right)^2}\, k^a, 
\end{align}
where $\xi^a=(1/3)\nabla_b h^{ba}=(\partial/\partial t)^a$ 
is the stationary Killing vector, 
$K_{ab}=f_{ac} f_b{}^c$ 
coincides with the Killing tensor in Eq.~\eqref{eq:KT}, 
and $C_{ab}=h_{ac} h_b{}^c$ is a conformal Killing tensor. 

The parallelly propagated tetrad 
$\{k^a, l^a, m^a, n^a\}$ along a null geodesic
becomes singular due to an infinite gravitational blue shift
if the geodesic approaches the horizon. 
Simultaneously, the divergence of $k^\phi$ means that a photon orbits the black hole 
infinite times during a finite interval of the affine parameter. 
In order to estimate the shear of a congruence of null geodesics 
during they wind finite times around the black hole, 
we should 
regularize the tetrad even on the horizon.  
Then, we introduce
\begin{align}
\label{tilde_k}
	\tilde{k}^a=\sqrt{\Delta}\, k^a,
\quad
	\tilde{l}^a=\frac{l^a}{\sqrt{\Delta}}. 
\end{align}
Note that $\tilde{k}^a$ and $\tilde{l}^a$ are 
parallelly propagated if we restrict the null geodesics 
to spherical photon orbits.

We evaluate the tetrad components of the Weyl tensor
$C_{\tilde{k}AB\tilde{k}}
:=C_{abcd} \tilde{k}^a(e_A)^b(e_B)^c\tilde{k}^d$ 
on the spherical photon orbits, 
where $(e_1)^a =m^a$ and $(e_2)^a =n^a$. 
Using Eqs.~\eqref{eq:bqsol1} and \eqref{eq:bqsol2}, we have 
\begin{align}
	&C_{\tilde{k} 11\tilde{k}}
	=-C_{\tilde{k} 22\tilde{k}}
		=\frac{12 r^3 \Delta^2 
		\left[ 5 (r^2-a^2 \cos^2\theta)^2-4 r^4 \right] }{(r-1)^2 \Sigma^5 },
\\
	&C_{\tilde{k} 12\tilde{k}}
	=C_{\tilde{k} 21\tilde{k}}
		=\frac{12 a r^2\Delta^2 \cos \theta 
		\left[ 5 (r^2-a^2\cos^2\theta)^2-4 a^4 \cos^4\theta \right]}
			{(r-1)^2 \Sigma^5}.
\end{align}
It is clear that these Weyl components are nonvanishing for the spherical 
photon orbits of the cardioid class, where $r > 1$.  
In contrast, for the spherical photon orbits of horizon class, 
taking the limit $\epsilon\to 0$ and $\delta \to 0$, we find 
\begin{align}
\label{eq:limCkmmk}
	&C_{\tilde{k} 11 \tilde{k}}=-C_{\tilde{k} 22\tilde{k}}
	\simeq\frac{12 \left(1-10 \cos^2\theta+5\cos^4\theta\right)}{(1+\cos^2\theta)^5}
	\left(\delta^2-4 \epsilon 
		\right)\to 0,
\\
\label{eq:limCkmnk}
	&C_{\tilde{k} 12 \tilde{k}}=C_{\tilde{k} 21 \tilde{k}}
	\simeq \frac{12\cos\theta \left(5-10\cos^2\theta+\cos^4\theta\right)}{(1+\cos^2\theta)^5} 
	\left(\delta^2-4 \epsilon \right)\to 0.
\end{align}
With the assumption that the congruence is twist-free, 
the evolution of the expansion, $\Theta$, 
and the shear, $\sigma_{AB}$, 
of a congruence of spherical photon orbits in Kerr spacetimes 
are determined by 
\begin{align}
	&\frac{\mathrm{d}}{\mathrm{d}\tilde\lambda} {\Theta}= -\frac{\Theta^2}{2}-\sigma^{AB}\sigma_{AB}, 
\\
	&\frac{\mathrm{d}}{\mathrm{d}\tilde\lambda} {\sigma}_{AB}=-\Theta \sigma_{AB}+C_{\tilde{k}AB\tilde{k}},
\end{align}
where 
$\tilde \lambda$ is a affine parameter on the spherical photon orbit. 
Then, Eqs.~\eqref{eq:limCkmmk} and \eqref{eq:limCkmnk} 
mean that the Weyl curvature does not produce the shear 
of a congruence of spherical photon orbits of the horizon class 
in near-extremal Kerr black holes. 
Hence, if the congruence has the initial conditions~$\Theta=0$ and $\sigma_{AB}=0$, 
then $\Theta$ and $\sigma_{AB}$ remains zero along the orbit.

In the extremal limit, $a \to 1$, 
we see, from Eqs.~\eqref{eq:eom} and \eqref{tilde_k}, 
that ${\tilde k}^a$ of a spherical photon orbit of the horizon class
approaches the horizon generator $\chi^a$, given by Eq.~\eqref{eq:chi}. 
On the other hand, 
the regularized outgoing principal null vector $(\Delta/2)N_+^a$ 
on the horizon 
is proportional to $\chi^a$. 
Although $(\Delta/2)N_+^a$ and ${\tilde k}^a$ are 
characterized by different values of the 
constants of motion $(b,q)$, both vectors approach 
the unique null vector on the event horizon 
$\chi^a$ in the limit $a \to 1$ and $r \to 1$. 
Therefore, we can understand that 
the spherical photon orbit of the horizon class is 
shear-free in this limit from the fact that 
the principal null geodesic is shear-free 
in the Kerr spacetime, which is classified in Petrov type D.

%%%%%%%%%%
\section{Summary and discussions}
%%%%%%%%%%

Vanishing shear of a congruence of 
the spherical photon orbits of the horizon class is 
interesting from the observational point of view. 
An image of a compact source through 
a congruence of light rays in the horizon class 
around a near-extremal Kerr black hole can keep its brightness 
even if the photons orbit around the black hole 
many times, 
and then, it would be observable. 
For extended sources, the spherical photon orbits correspond to the bright border of 
the black hole shadow~\cite{Akiyama:2019cqa, Akiyama:2019brx, 
Akiyama:2019sww,Akiyama:2019bqs, Akiyama:2019fyp, Akiyama:2019eap, Bardeen:1973tla, Young:1976zz, Falcke:1999pj, 
Doeleman:2008qh, Doeleman:2012zc}.  
In this case, photon orbits of the horizon class and the ones of the cardioid class 
would make high contrast of brightness of the shadow border for 
an equatorial observer. 
It is an interesting and important next work to clarify the relation quantitatively 
between the contrast and spin parameter of the black hole. 

%\red{\sout{If the horizon class appears in a rapidly rotating black hole, the shadow edge can be bright even for nonspreading light sources.}}

%\red{\sout{
%Furthermore, 
%it takes a large amount of time a photon travels 
%from a distant source 
%through 
%a near-horizon region to a distant observer, and therefore, 
%we can observe the early stage of the Universe 
%by the gravitational lensing of a 
%near-extremal Kerr black hole. 
%The spherical photon orbits in 
%near-extremal Kerr black hole would be a photon storage 
%in nature. }}

\begin{acknowledgments}
The authors thank 
T.~Harada, 
S.~Kinoshita, 
T.~Koike, 
K.~Nakao, 
J.~M.~M.~Senovilla,
R.~Takahashi,
and Y.~Yasui for useful comments. 
This work was supported by Grant-in-Aid for Early-Career Scientists [JSPS KAKENHI Grant No.~JP19K14715] (T.I.) and Grant-in-Aid for Scientific Research(C) [JSPS KAKENHI Grant No.~JP16K05358] (H.I.) from the Japan Society for the Promotion of Science and 
the MEXT-Supported Program for the Strategic Research Foundation at Private Universities, 2014--2017 (S1411024) from the Ministry of Education, Culture, Sports, Science and Technology (T.I.).
%This work was supported by JSPS KAKENHI Grants No.~JP16K05358~(H.I.), 
%No.~JP19K14715~(T.I.)
%and the MEXT-Supported Program for the Strategic Research Foundation at Private Universities, 
%2014--2017~(S1411024)~(T.I.). 
 \end{acknowledgments}

\end{document}